%% file: main.tex
\documentclass[sigconf]{acmart}
\AtBeginDocument{%
  }

\usepackage[most]{tcolorbox}
\newtcolorbox{summarybox}[2][]{%
  colback=gray!5,
  colframe=gray!40,
  fonttitle=\bfseries,
  title=#2,
  sharp corners,
  boxrule=0.5pt,
  left=6pt,
  right=6pt,
  top=6pt,
  bottom=6pt,
  #1
}

\usepackage[dvipsnames]{xcolor} 
\usepackage{minted}
\usepackage{enumitem}
\usepackage{mdframed}
\mdfsetup{font=\normalsize}
\usepackage{enumitem}
\usepackage{amsmath}
\usepackage{balance}
\usepackage{subcaption}
\usepackage{graphicx}
\usepackage{booktabs}
\usepackage{relsize}
\usepackage{colortbl}
\usepackage{tikz}
\usepackage{pgf}
\usepackage{pbox}
\usepackage{rotating}
\usepackage{multirow}
\usepackage{lipsum}
\usepackage{tablefootnote}
\usepackage{cleveref}
\crefformat{footnote}{#2\footnotemark[#1]#3}
\usepackage{soul}

\usepackage{url}
\usepackage{algpseudocode}
\usepackage{listings}
\usepackage{float}

\acmYear{2026}\copyrightyear{2026}
\setcopyright{cc}
\setcctype[4.0]{by}
\acmConference[FSE Companion '26]{34th ACM Joint European Software Engineering Conference and Symposium on the Foundations of Software Engineering}{July 5--9, 2026}{Montreal, QC, Canada}
\acmBooktitle{34th ACM Joint European Software Engineering Conference and Symposium on the Foundations of Software Engineering (FSE Companion '26), July 5--9, 2026, Montreal, QC, Canada}
\acmDOI{10.1145/3803437.3805236}
\acmISBN{979-8-4007-2636-1/2026/07}

\newcommand{\startlist}{\begin{list}{\labelitemi}{\leftmargin=1em}\setlength{\itemsep}{-1mm}}
\newcommand{\stoplist}{\end{list}}

\newcommand{\rqone}{To what degree do the assessment strategies in HalluJudge effectively detect hallucinations in code review comments?}

\newcommand{\rqtwo}{How efficient is HalluJudge in detecting code review hallucination?}

\newcommand{\rqthree}{To what degree do HalluJudge’s hallucination judgments align with developers’ preferences in practice?}

\begin{document}


\title{HalluJudge: A Reference-Free Hallucination Detection for Context Misalignment in Code Review Automation}


\author{Kla Tantithamthavorn}
\affiliation{%
  \institution{Monash University}
  \country{Australia.}
}
\email{chakkrit@monash.edu}

\author{Hong Yi Lin}
\affiliation{%
  \institution{The University of Melbourne}
  \country{Australia.}
}
\email{holin2@student.unimelb.edu.au}

\author{Patanamon Thongtanunam}
\affiliation{%
  \institution{The University of Melbourne}
  \country{Australia.}
}
\email{patanamon.t@unimelb.edu.au}

\author{Wachiraphan Charoenwet}
\affiliation{%
  \institution{The University of Melbourne}
  \country{Australia.}
}
\email{wcharoenwet@student.unimelb.edu.au}

\author{Minwoo Jeong}
\affiliation{%
  \institution{Atlassian}
  \country{USA.}
}
\email{mjeong@atlassian.com}

\author{Ming Wu}
\affiliation{%
  \institution{Atlassian}
  \country{USA.}
}
\email{mwu2@atlassian.com}

\renewcommand{\shortauthors}{Tantithamthavorn et al.}

\begin{abstract}
Large Language models (LLMs) have shown strong capabilities in code review automation, such as review comment generation, yet they suffer from hallucinations---\emph{where the generated review comments are ungrounded in the actual code}---poses a significant challenge to the adoption of LLMs in code review workflows. 
To address this, we explore effective and scalable methods 
for a hallucination detection in LLM-generated code review comments without the reference. 
In this work, we design HalluJudge that aims to assess the grounding of generated review comments based on the context alignment.
HalluJudge includes four key strategies ranging from direct assessment to structured multi-branch reasoning (e.g., Tree-of-Thoughts). 
We conduct a comprehensive evaluation of these assessment strategies across Atlassian's enterprise-scale software projects to examine the effectiveness and cost-efficiency of HalluJudge.
Furthermore, we analyze the alignment between HalluJudge's judgment and developer preference of the actual LLM-generated code review comments in the real-world production. 
Our results show that the hallucination assessment in HalluJudge is cost-effective with an F1 score of 0.85 and an average cost of \$0.009. 
On average, 67\% of the HalluJudge assessments are aligned with the developer preference of the actual LLM-generated review comments in the online production.
Our results suggest that HalluJudge can serve as a practical
safeguard to reduce developers’ exposure to hallucinated comments,
fostering trust in AI-assisted code reviews.
\end{abstract}



\maketitle

\section{Introduction}



Large Language Models (LLMs) have rapidly transformed the landscape of software engineering, particularly in automating code review processes~\cite{tufano2021empirical,li2023chatgpt,she2023pitfalls}.
By generating natural language feedback on code changes, these models promise to enhance developer productivity and code quality~\cite{code_review_survey2022}. However, a critical challenge remains: LLMs are prone to generating \textbf{hallucinated code review comments}---i.e., \emph{feedback that is ungrounded, irrelevant, or misaligned with the actual code changes}~\cite{ji2023survey}. Recent studies have shown that such hallucinations are prevalent in LLM-based software engineering tasks~\cite{pearce2025asleep,nijkamp2023codegen}. In the context of code review, hallucinated comments can erode developer trust, introduce confusion, and ultimately hinder the adoption of automated code review tools in real-world software development environments~\cite{fan2023large}.

Detecting code review hallucinations is a non-trivial task, especially in the absence of explicit answers or gold-standard annotations.
Existing evaluation methods often rely on reference-based metrics (e.g., BLEU Score~\cite{bleu}, Exact Match) or manual inspection, both of which are limited in scalability and generalizability.
For example, reference-based metrics assume the availability of a single review comment (i.e., an actual human-written comment) and primarily measure surface-level lexical overlap, failing to capture whether a comment is semantically grounded in the reviewed code or aligned with its surrounding context~\cite{ling2015not}.
Recent work also shows that traditional reference-free metrics in natural language processing (e.g., feature attribution, uncertainty metrics) perform poorly at detecting hallucinations in code reviews~\cite{liu2025hallucinations}.
Manual evaluation, while more reliable, is expensive, subjective, and difficult to reproduce at scale~\cite{howcroft2020twenty}.

While recent research has focused on addressing hallucinations in code generation produced by LLMs~\cite{tian2025codehalu,maharaj2025etf}, review comment generation presents a fundamentally different challenge. 
Unlike code generation tasks, review comments require LLMs to produce natural language critique grounded in code changes.
This makes hallucinations harder to detect because they cannot be validated through compilation or runtime checks. 
Instead, they involve semantic misalignment between the comment and the actual code change, which demands new evaluation strategies beyond traditional program analysis.


In this paper, we present HalluJudge, a suite of reference-free hallucination judgment approaches designed to assess the grounding of generated review comments.
Our HalluJudge systematically assesses the alignment between code review comments and code diffs by decomposing the reasoning process into a structured, multi-step evaluation, inspired by expert human reviewers. 
Specifically, HalluJudge includes four key strategies ranging from direct assessment to structured multi-branch reasoning (e.g., Tree-of-Thoughts), enabling a comprehensive exploration of reference-free hallucination detection.
These approaches enable robust detection of hallucinations without the need for references, making it suitable for large-scale, real-world deployments of automated code review tools.

To explore effective hallucination assessments in HalluJudge, we conduct an extensive assessment  using code review data from Atlassian's enterprise software development projects. 
We further validate our approach by examining the alignment with developer preference (i.e., thumbs up) of LLM-generated code review comments in production environments. 
We answer the following RQs:

\begin{enumerate}[label=\textbf{RQ\arabic*)}]

\item {\bf \rqone}\\
The direct and tree of thought strategies in HalluJudge can effectively detect hallucinations in code review comments, achieving a precision, recall, and F1 score of 0.85. Among the four assessment strategies, the tree of thought approach delivers the highest scores across all three metrics.

\item {\bf \rqtwo}\\
The direct assessment in HalluJudge is cost-effective in terms of the token used and monetary cost, whereas the tree of thought strategy achieves the best performance but requires the highest cost.

\item {\bf \rqthree}\\
HalluJudge assessment is well aligned with the developer preference signals in the online production with an average consistency and coverage score up to 0.67.
\end{enumerate}

Our results show that HalluJudge can cost-effectively detect hallucinated review comments.
Integrating HalluJudge as a practical safeguard into the automated code review could reduce developers’ exposure to hallucinated comments, addressing a key barrier to the adoption of automated code reviewers. 
By preventing plausible yet incorrect feedback, we believe it will enhance perceived reliability and foster long-term trust in AI-assisted code reviews.


\textbf{Novelty.} 
To the best of our knowledge, we are the first to:

\begin{itemize}[leftmargin=1.5em, itemsep=0pt]
    \item Introduce a reference-free hallucination detection for detecting context-misaligned code review comments.
    \item Extensively evaluate various assessment strategies for detecting hallucinated review comments based on the enterprise-scale software projects at Atlassian.
    \item Quantify the alignment between hallucination judgment of LLMs and developers' preferences of actual generated review comments in the online production environment.
\end{itemize}



\section{Background \& Related Work}

In this section, we discuss background and related work to highlight the research gaps with respect to the large language models (LLMs) for automated code review, and hallucination in SE. 

\subsection{Automated Code Review}

To improve software quality, reduce reviewer burden, and accelerate the code review process, significant research efforts have focused on automating code review—particularly the generation of review comments~\cite{hong2022commentfinder,li2022automating}.
Recent advances in large language models (LLMs) have further enabled the automation to generate fluent, human-like comments from code changes~\cite{lu2023llama,tufano2022using,tufano2021towards,lin2024leveraging,Tantithamthavorn2026RovoDevCR}.
Unlike traditional automated review tools that primarily detect defects or enforce coding standards through static analysis or rule-based checks~\cite{wattanakriengkrai2020predicting,charoenwet2024empirical}, review comment generation aims to produce natural-language feedback similar to human-written comments for code changes in a pull request (PR). 
This task is critical because well-crafted comments not only identify issues but also explain their rationale and suggest actionable improvements, thereby enhancing developer understanding and collaboration.



Although LLM-based approaches show considerable promise for automating code review, generating high-quality review comments remains a significant challenge~\cite{liu2025too}.
Recent studies report that fewer than 10\% of benchmark comments meet established quality criteria~\cite{lu2025deepcrceval}.
Empirical evidence from real-world deployments further reveals that, although LLMs can generate diverse types of concerns, many comments remain unresolved due to issues of clarity, relevance, and simplicity~\cite{goldman2025types}.
Moreover, Liu et al.~\cite{liu2025hallucinations} highlight that even fine-tuned LLMs frequently produce hallucinations—introducing misaligned or fabricated details not grounded in the code changes.

Evaluating the quality of LLM-generated outputs remains a major challenge in software engineering research. 
Many existing approaches rely on reference-based metrics such as BLEU or Exact Match, which compare generated text against a predefined ground-truth output~\cite{bleu}. While effective for tasks with a single correct answer, these metrics are not suitable for code review, where multiple valid comments may exist and lexical similarity does not necessarily reflect semantic correctness or contextual grounding~\cite{ling2015not}.
Moreover, prior work reveal that these metrics often fail to capture critical dimensions such as correctness, and contextual grounding~\cite{lin2024leveraging,yang2023evacrc}. This highlights the urgent need for  hallucination detection to ensure that generated review comments are accurate and relevant, thereby increasing trustworthiness and promoting adoption in practice.


\subsection{Hallucination in LLM-Based Software Engineering}

Hallucination refers to the phenomenon where LLMs generate content that is fluent and plausible but not grounded in the input or underlying facts~\cite{ji2023survey}. In software engineering tasks, hallucinations may manifest as incorrect program explanations, fabricated APIs, or unfounded bug reports, which can be particularly harmful given the precision required in code-related activities~\cite{pearce2025asleep,nijkamp2023codegen}.

Prior work has proposed various taxonomies of hallucination in natural language generation, often distinguishing between \emph{intrinsic} errors (e.g., contradictions with the input) and \emph{extrinsic} errors (e.g., unsupported claims)~\cite{ji2023survey}. In the context of code review, however, hallucination frequently appears as \emph{context misalignment}, a subtype of intrinsic error, where a comment is syntactically valid and stylistically appropriate but semantically inconsistent with the actual code changes or their surrounding context~\cite{liu2025hallucinations}. 
For example, Figure~\ref{fig:hallucination-example} shows that a PR only changes how the service is invoked (synchronous call $\rightarrow$ asynchronous call).
However, the LLM-generated review comment incorrectly claims that the change introduces an SQL injection vulnerability.
This is considered as a \emph{hallucinated} comment because it is not grounded in the code change as it does not modify any SQL queries or introduce user-controlled query strings.
This distinction is critical, as misaligned review comments may not be obviously incorrect to developers and can therefore undermine trust in automated review tools over time~\cite{fan2023large}.

Extensive research has primarily focused on hallucination detection in code generation tasks. For example, CodeHalu identifies hallucinations in generated code by detecting common error patterns during execution~\cite{tian2025codehalu}, while ETF leverages program analysis techniques to detect hallucinations in code summarization~\cite{maharaj2025etf}. These approaches mainly target compilable code implementations, where hallucinations can often be detected through static analysis or execution-based verification. 
However,these approaches are not applicable to review comment generation, where LLMs produce natural language rather than executable code.
Recent work also shows that traditional reference-free metrics from natural language processing (e.g., feature attribution, uncertainty-based measures) perform poorly at detecting hallucinations in code review comments~\cite{liu2025hallucinations}. Manual evaluation, although more reliable, is costly, subjective, and difficult to reproduce at scale~\cite{howcroft2020twenty}. This highlights the need for a novel method for reference-free hallucination detection in the review comment generation.

\begin{figure}[t]
\centering
\small
\begin{tabular}{p{0.95\linewidth}}
\hline
\textbf{Diff (Java):} \\
\texttt{@@ -42,7 +42,9 @@ public Result handle(Request req) \{} \\
\texttt{-    User u = userService.getUser(req.getUserId());} \\
\texttt{-    return ok(render(u));} \\
\texttt{+    CompletableFuture<User> f = userService.getUserAsync(req.getUserId());} \\
\texttt{+    return f.thenApply(u -> ok(render(u)))} \\
\texttt{+            .exceptionally(e -> internalServerError());} \\
\hline
\textbf{LLM Code Review Comment (Hallucinated):} \\
\emph{``This change looks vulnerable to SQL injection since \texttt{getUserId()} is used directly.
You should sanitize inputs before constructing the query.''} \\
\hline
\textbf{Why this is hallucination:} \\
The diff does not construct or modify any SQL query, does not introduce string concatenation, and does not change database access logic.
The comment is plausible in general, but it is \emph{unsupported} and \emph{misaligned} with the actual code changes. \\
\hline
\end{tabular}
\caption{An example of a hallucinated code review comment  (i.e., claims about an SQL injection risk) that is context-misaligned with the code diff (i.e., no area of code related to SQL or user-controlled queries).
}
\label{fig:hallucination-example}
\end{figure}

\section{HalluJudge: Hallucination Judges for Detecting the Context Misalignment in Code Review}

In this section, we present HalluJudge, i.e., hallucination judges for detecting the context misalignment in code review.
\subsection{Overview}
HalluJudge is a context-aware hallucination detection framework designed to evaluate how well LLM-generated review comments are grounded in the underlying code changes. 
This is different from the conventional \emph{LLM-as-a-Judge} approaches which primarily assess surface-level plausibility or score the generated outputs against a set of criteria without considering the context ~\cite{gu2024surveyllmasajudge,zheng2023judging,dubois2024lengthcontrolled}.

HalluJudge formulates the hallucination detection as a problem of \emph{context alignment}: the generated review comment is supported by, traceable to, and consistent with the factual elements of the code changes.
Hence, HalluJudge explicitly grounds judgment in the review context, analyzing the comment and the claims it makes, then determine the alignment of those claims with the code changes.
This formulation is closely related to evidence-grounded evaluation paradigms that decompose outputs into atomic claims and verify them against a reference context~\cite{min2023factscore,es2023ragas}. 

\subsection{Definitions}
\label{sec:def}In this paper, a \emph{code diff} ($\mathbb{D}$) denotes the code changes in a pull request (PR), which serves as the primary \emph{context} for automated code review.
A \emph{review comment} ($\mathbb{C}$) denotes a comment generated by an LLM based on the code diff $\mathbb{D}$.
Each review comment consists of a set of \emph{claims} ($\{c_1,c_2,...,c_n\}$) that pinpoint potential issues or provide suggestions for the code diff $\mathbb{D}$. 

Given that a code diff$(\mathbb{D})$  represents the factual elements and the generated review comment $\mathbb{C}$ contains the claimed concerns or suggestions by the LLM, we assess whether each claim ($c_i$) is supported by the underlying factual elements.  
Hence, we define a grounding function $\mathbb{G}$, which maps each claim to a binary judgment indicating whether it is directly supported by at least one factual element in $(\mathbb{D})$. 
Formally, 
\[
\mathbb{G}(c_i) =
\begin{cases}
1, & \text{if } c_i \text{ is directly supported by } \exists d_i \in \mathbb{D} \\
0, & \text{otherwise}.
\end{cases}
\]
A claim is considered grounded if the evidence in the code diff ($\mathbb{D}$) fully entails the claim; partial correctness is insufficient and results in the claim being treated as ungrounded.
A code review is considered \textbf{\emph{hallucinated}} when the review comment contains at least one ungrounded claim.
Formally,
\[
\text{Hallucination}(\mathbb{C}, \mathbb{D}) =
\begin{cases}
1 & \text{if } \exists c_i \in \mathbb{C} \text{ such that } \mathbb{G}(c_i) = 0 \\
0 & \text{otherwise}
\end{cases}
\]

\subsection{Instructions for Hallucination Judgment}
\label{sec:instructions}
To operationalize hallucination judgment defined in Section \ref{sec:def}, we formalize the task as assessing the context misalignment between the claims in a review comment and the facts presented in the code diff.
We then define the context misalignment based on the our grounding function $\mathbb{G}$.
As a review comment may contain a mix of both grounded and ungrounded claims, we define five levels of context alignment using a scoring-based evaluation. 
The system prompt used for this task is shown below.
Finally, we define the user prompt contains the generated review comment $\mathbb{C}$ and its corresponding code diff $\mathbb{D}$.

\begin{mdframed}[backgroundcolor=green!5, linecolor=green!60!black, linewidth=1pt, roundcorner=5pt, innertopmargin=5pt, innerbottommargin=5pt, skipabove=1em,  frametitle={\textbf{HalluJudge's System Prompt}},
  frametitlealignment=\center,
  frametitlerule=true,frametitleaboveskip=0.6em,
    frametitlebelowskip=0.6em]

\noindent\textbf{[Task Description]}\\
You are an LLM-Judge in software engineering tasked with directly assessing whether a review comment (the "claims") is context-misaligned with a given code diff (the "facts").\\
\noindent\textbf{[Definition]} \\
Context misalignment occurs when an LLM-generated code review comment contains information, \\claims, or suggestions that:
\begin{itemize}
\setlength{\leftmargini}{0.1em}
\item Are not supported by, or cannot be traced to, the relevant context (e.g., code diff, or input text).
\item Contradict the facts or intent present in the context.
\item Reference information that is missing, irrelevant, or outside the scope of the provided context.
\end{itemize}

\noindent\textbf{[Scoring Guide]}

\noindent0) Fully aligned; all claims are directly supported and consist ... 

\noindent1) Mostly aligned; most claims supported, minor inconsiste ...

\noindent2) Partially aligned; some claims are supported, others are ...

\noindent3) Mostly misaligned; very little support from the code diff ... 

\noindent4) Completely misaligned; claims are unsupported or contr ...\\
\noindent\textbf{[Output]} \\
Provide your evaluation ... using a scores-based evaluation of context misalignment levels (0–4), along with a concise explanation referencing specific evidence from the code ...

\noindent \textbf{[Example Output]}\\ \{ "answer": 0-4, "explanation": "..."\}

\end{mdframed}

\begin{mdframed}[backgroundcolor=green!5, linecolor=green!60!black, linewidth=1pt, roundcorner=10pt, innertopmargin=10pt, innerbottommargin=10pt, skipabove=1em,  frametitle={\textbf{User Prompt}},
  frametitlealignment=\center,
  frametitlerule=true,frametitleaboveskip=0.6em,
    frametitlebelowskip=0.6em]
    
\textbf{[Diff]...[/Diff]}\\
\textbf{[LLM Code Review]...[/LLM Code Review]}\\
Please evaluate whether this review comment is context-aligned or
misaligned with the given code diff. Only provide the valid
JSON format as specified in the system prompt.
\end{mdframed}

\subsection{Diverse Assessment Strategies}
To assess how well a review comment ($\mathbb{C}$) aligns with its corresponding code diff ($\mathbb{D}$), we explore several assessment strategies that leverage different levels of reasoning capability.
Specifically, we examine four strategies: (1) direct assessment, (2) few-shot direct assessment, (3) multi-step reasoning, and (4) tree‑of‑thoughts reasoning. These strategies provide a comprehensive investigation of effective methods for assessing context alignment in this task. Below, we briefly describe each assessment strategy.



\subsubsection{Direct assessment} Following the prior work~\cite{zhou2025llm}, this strategy leverages the zero-shot capabilities of LLMs by prompting the model to directly make a judgment  without using a task-specific or structured reasoning procedure. 
This strategy demonstrates the model's out-of-the-box reasoning capability without human intervention, while also offer low LLM usage costs~\cite{brown2020language, wei2023larger}.
For this strategy, HalluJudge directly uses the system and user prompts defined in Section \ref{sec:instructions} to assess the context misalignment of a review comment.

\begin{mdframed}[backgroundcolor=green!5, linecolor=green!60!black, linewidth=1pt, roundcorner=5pt, innertopmargin=5pt, innerbottommargin=5pt, skipabove=1em,  frametitle={\textbf{System Prompt of Few-Shot Strategy}},
  frametitlealignment=\center,
  frametitlerule=true,frametitleaboveskip=0.6em,
    frametitlebelowskip=0.6em]

\noindent\textbf{[Task Description]}
...\\
\noindent\textbf{[Definition]}
...\\
\noindent\textbf{[Scoring Guide]}
...\\
\noindent\textbf{[Examples]}

\noindent\{Fully Aligned Example\}

\noindent\{Mostly Aligned Example\}

\noindent\{Partially Aligned Example\}

\noindent\{Mostly Misaligned Example\}

\noindent\{Completely Misaligned Example\}

\noindent[\textbf{Output Format]}
...\\
\noindent \textbf{[Example Output]} ...
\end{mdframed}

\subsubsection{Few-shot assessment}
Few-shot prompting has been shown to substantially improve performance on complex reasoning and classification tasks by providing in-context demonstrations~\cite{brown2020language, dong2022survey}.
This strategy helps the LLM more clearly understand the decision boundaries needed to infer different levels of context misalignment based on the claims in a review comment.
Hence, we explore this assessment strategy by providing five examples, corresponding to each of the scoring levels.
We include these labeled examples in the system prompt to guide the LLM’s judgment as shown above.

\begin{mdframed}[backgroundcolor=green!5, linecolor=green!60!black, linewidth=1pt, roundcorner=10pt, innertopmargin=10pt, innerbottommargin=10pt, skipabove=1em,  frametitle={\textbf{A Multi-Steps Reasoning Assessment}},
  frametitlealignment=\center,
  frametitlerule=true,frametitleaboveskip=0.6em,
    frametitlebelowskip=0.6em]

\noindent\textbf{[Task Description]}
...\\
\noindent\textbf{[Definition]}
...\\
\noindent\textbf{[Scoring Guide]}
...\\ 
\noindent\textbf{[Multi-Step Reasoning Evaluation Process]}

1. Comprehension: Carefully read the code diff and the LLM-generated code review comment ... 

2. Traceability Check: For each claim or suggestion in the review comment, identify the specific lines or sections in the code diff that support or contradict it ...

3. Support Assessment: Determine whether the evidence from the code diff fully supports ...

4. Conflict and Relevance Analysis: Check if any part of the review comment conflicts with the actual code changes ... 

5. Context Completeness Consideration: Consider whether the code diff provides enough context ...

6. Final Judgement: Based on the above steps, decide if the review comment is context aligned ...

\noindent \textbf{[Output Format]}
...\\
\noindent \textbf{[Example Output]} ...
\end{mdframed}

\subsubsection{Multi-steps reasoning} This strategy is inspired by chain-of-thought prompting, which has been demonstrated to enhance reasoning accuracy and interpretability on challenging tasks~\cite{wei2022chain, kojima2022large}.
Multi-step reasoning decomposes the judgment process into intermediate reasoning steps and explicitly instructs an LLM to follow the procedure before producing a final decision. 
In HalluJudge, we define five steps of assessment which conceptually centers on determining whether each claim in a review comment can be faithfully traced back to the evidence provided in a code diff. It involves understanding both the diff and the comment, checking whether each claim is supported or contradicted by the code changes, identifying any irrelevant or out‑of‑scope references, and considering whether the diff itself provides sufficient context. By examining the grounding, support, and relevance of each claim, the process  helps a judgment of whether the review comment is context‑aligned or context‑misaligned.
We include this procedure in the system prompt to guide the LLM’s judgment as shown above.



\begin{mdframed}[backgroundcolor=green!5, linecolor=green!60!black, linewidth=1pt, roundcorner=10pt, innertopmargin=10pt, innerbottommargin=10pt, skipabove=1em,  frametitle={\textbf{Tree-of-Thoughts Assesment}},
  frametitlealignment=\center,
  frametitlerule=true,frametitleaboveskip=0.6em,
    frametitlebelowskip=0.6em]

\noindent\textbf{[Task Description]}
...\\
\noindent\textbf{[Definition]}
...\\
\noindent\textbf{[Scoring Guide]}
...\\ 
\textbf{[Tree of Thoughts Process]} You will explore multiple reasoning paths simultaneously before making a final judgment. \\
\textbf{Branch A: Alignment Hypothesis.}  
Assume the comment \textbf{IS} context-aligned. Find evidence: What specific diff elements support each claim in the comment? How do the comment's suggestions relate to visible changes? ...
\\
\textbf{Branch B: Misalignment Hypothesis.} 
Assume the comment is \textbf{NOT} context-aligned. Find evidence: What claims in the comment lack support from the diff? What contradictions exist between comment and actual changes? ...
\\
\textbf{Branch C: Evidence Mapping.} 
Map each sentence/claim in the comment to diff elements: Which lines in the diff correspond to each comment claim? ...
\\
\textbf{Branch D: Context Boundaries.} 
Evaluate what the comment should or should not reference: What would require external context not provided? Is the comment's scope appropriate for the given diff? ...
\\
\textbf{[Synthesis Process]}
\begin{itemize}[leftmargin=1em]
    \item Evidence Weight: Which branch found stronger, more specific evidence?
    \item Contradiction Analysis: Do branches conflict? How do you resolve them?
    \item Confidence Assessment: Which reasoning path feels most reliable?
    \item Final Judgment: Based on the strongest evidence path
\end{itemize}
\noindent \textbf{[Output Format]}
...\\
\noindent \textbf{[Example Output]} ...

\end{mdframed}

\subsubsection{Tree of Thoughts}
Extending the multi-steps reasoning, the Tree of Thoughts (ToT) explicitly instructs an LLM to explore multiple reasoning paths in parallel and evaluate alternative hypotheses before reaching a conclusion~\cite{yao2024tree}. Instead of following a single linear reasoning chain, the model branches into several possible lines of thought, reflects on their validity, and selects the most coherent path. This structured exploration supports more robust judgment on complex tasks.



In HalluJudge, we define four complementary reasoning branches to assess how well a review comment is grounded in its corresponding code diff. These branches target four key dimensions: alignment, misalignment, traceability, and contextual scope.
The \textbf{alignment hypothesis}
 attempts to identify supporting factual element (i.e., the diff elements) for the corresponding claim.
The \textbf{misalignment hypothesis}  attempts to search for claims that (i) contradict with the diff, (ii) are not supported by  the code diff, or (iii) rely on external assumptions absent from the review context. 
The \textbf{claim-to-diff traceability} enforces high-fidelity grounding by constructing explicit trace links.
This produces interpretable artifacts suitable for auditing LLM-based judges~\citep{zheng2023judging}.
The \textbf{context boundary}
verifies that claims remain within the intended scope of the change and repository constraints. 
Even factually grounded statements may constitute hallucinations if they introduce scope creep (e.g., speculative refactors or unrelated security concerns). 
Finally, we instruct an LLM to analyze and synthesize the branches and make the judgment based on the strongest evidence path.
We include this procedure in the system
prompt to guide the LLM’s judgment as shown above.

\section{Study Design: An Industrial Case Study at Atlassian}
\label{sec:experimental-setup}

In this Section, we present the research questions, datasets, and research methodology.

\subsection{Research Questions}
\label{sec:research-questions}

The goal of our experimental evaluation is to assess the effectiveness of HalluJudge in detecting context misalignment in LLM-generated code review comments, serving as one of the quality safeguards for automated code review in practice. 
Therefore, we set out to investigate the following three research questions (RQs):

\begin{enumerate}[label=\textbf{RQ\arabic*}]
    \item \textbf{\rqone}
    In RQ1, we aim to systematically explore diverse assessment strategies of HalluJudge in detecting hallucination, i.e., context misalignment, in LLM-generated code review comments. To achieve this, we evaluate the effectiveness to determine which are most suitable for analysing the claims made in code review comments and accurately judging their alignment with the surrounding code context.

    \item \textbf{\rqtwo}
     HalluJudge aims to help automated code review to safeguard the generated review comments before they reach developers. However, automated code reviewers can generate a large number of review comments in industrial settings, making the computational cost of applying HalluJudge a practical consideration. Therefore, in RQ2, we therefore investigate the efficiency of HalluJudge by measuring token usage and monetary cost when detecting code review hallucinations.

    \item \textbf{\rqthree} \\
    The practical value of hallucination detection ultimately depends on how well it reflects developer expectations and real-world usage. Implicit signals from developers would provide ecologically valid indicators of comment quality~\cite{Tantithamthavorn2026RovoDevCR}. Therefore, in RQ3, we investigate whether HalluJudge’s judgments is aligned with the code review comments preferred by developers in production settings. Strong alignment would indicate that HalluJudge can serve not only as an evaluation tool but also as a practical safeguard for deploying LLM‑based code review systems at scale.

\end{enumerate}

\subsection{Datasets}
\label{sec:dataset}

In this work, we used LLM-generated code review comments generated by Atlassian's \emph{RovoDev Code Reviewer}~\cite{Tantithamthavorn2026RovoDevCR}\footnote{https://www.atlassian.com/software/rovo-dev} that is being used internally on the proprietary enterprise software systems.
This is to avoid the data leakage of software artifacts used for LLM pre/post-training and other sensitive development artifacts, ensuring that all data remains within controlled enterprise environments.
RovoDev Code Reviewer~\cite{Tantithamthavorn2026RovoDevCR}, an LLM-based code review assistant which supports various underlying LLMs.
RovoDev Code Reviewer has been used by over 4,000 Atlassian software engineers for more than one year.
So far, RovoDev has generated more than 40{,}000 code review comments/month in production over one year, assisting developers by identifying potential issues, suggesting improvements, and highlighting risks during pull request reviews~\cite{goldman2025types}.
The system operates across 10 programming languages, 2,500 repositories, providing a diverse and realistic evaluation setting.

To answer our RQs, we require a dataset of LLM‑generated review comments paired with reliable ground‑truth signals indicating whether each comment is hallucinated or not. To this end, we construct two complementary evaluation datasets: \textbf{human annotations as ground truth}, and \textbf{
developer preference as practical quality signals}. 
The human‑annotated dataset is used to evaluate the effectiveness and cost‑efficiency of HalluJudge in detecting hallucinations, i.e., context misalignment, in code review comments (RQ1 and RQ2). 
The developer‑preference dataset is used to examine how well HalluJudge’s judgments align with the comments preferred by developers in real‑world usage (RQ3).
Below, we describe the data construction process for each dataset.

\subsubsection{Human Annotation as Ground-Truth   (RQ1 \& RQ2)}
\label{sec:human-annotated-dataset}
Given a large number of PRs and review comments in Atlassian's internal projects, manual annotation at scale is infeasible. 
To construct a manageable size of the dataset, we sampled merged PRs from 14 recent internal projects. 
To ensure balanced representation, we performed joint stratified sampling based on five commonly used programming languages (i.e., Java, Python, JavaScript, TypeScript, Kotlin), and two types of code changes (i.e., Add, Modify).
We sampled a total of 97 PRs, which provides a 95\% confidence interval with a maximum margin of error of ±10 percentage points for the population proportion.
The sampled PRs contain an average of 52 changed lines with ±26 of standard deviation ($\sigma$), across an average of 4 ($\sigma$=±3) files and 3 (±2) directories.
To ensure diversity in the generated code reviews, we employed three different commercial LLMs (i.e., Claude-Sonnet-4, Qwen3-Coder, and GPT‑5) as an engine in RovoDev Code Reviewer.
This process produced a total of 143 generated review comments.


\input{tables/annotation_dataset}

\paragraph{Hallucination Annotation.}
To determine if the generated reviews are hallucinated, three experienced software engineering researchers were engaged in the annotation process.
This involved manually checking each LLM-generated review comment against the attached code diff to determine if the comment is hallucinated (i.e., contextually misaligned) or non-hallucinated as defined in Section \ref{sec:def}.
The annotation followed a predefined workflow that requires reviewers to verify the comment's factual grounding, traceability to the code diff, and adherence to the change intent.
Two annotators independently labeled all 143 generated comments across three rounds of 50, 50, and 43 samples.
Between each round, the two annotators discussed and resolved any conflicting labels with the third annotator acting as a tie breaker.
The Cohen's Kappa ($\kappa$) for the three rounds were 0.78, 0.81, and 0.84.
This dataset is used exclusively to answer RQ1 and RQ2.
Table~\ref{table:offline_dataset} shows the statistical distributions of the human annotation dataset.

\subsubsection{Developer Preference as Signals (RQ3)}
\label{sec:developer-preference-dataset}

To evaluate HalluJudge at scale and under real-world settings, we leverage developer feedback collected directly from the online production environment.
Bitbucket, Atlassian's pull request platform, allows developers to provide thumbs‑up and thumbs‑down reactions.
Hence, we use this reaction as explicit developer preference on the LLM-generated review comments.

We collect 2{,}000 LLM-generated code review comments generated by RovoDev Code Reviewer over a three-month period.
We treat a thumbs-up as a signal that the comment is acceptable by developers hence the comments should be contextually aligned and non-hallucinated, while a thumbs-down implies the low quality that may potentially be hallucinated.
Comments without developer preferences are excluded from analysis.

In total, we obtain 557 generated code review comments with developer feedback.
This is consistent with prior studies on implicit developer feedback~\cite{frommgen2024resolving} where only approximately 5\% of comments receive explicit feedback.
Out of the 557 generated review comments, 370 (65\%) received thumb-up reactions.
This dataset enables a more practical evaluation of how well HalluJudge's judgments align with developer preferences in real-world software development workflows (RQ3).

\subsection{Experimental Setup}

To answer our RQs, we conduct a systematic evaluation using the dataset described in Section~\ref{sec:dataset}. 
Each dataset instance consists of a code diff paired with its corresponding code review comment. 
For every instance, we apply our \textit{HalluJudge} to assess whether the review comment are hallucinated. 

We perform experiments using two LLMs, i.e., Gemini 3 and GPT 5.1, across four assessment strategies. This results in a total of 8 experimental configurations, enabling a comprehensive examination of both the effectiveness and efficiency of hallucination judgment.
Below, we describe the LLMs, analysis approaches and evaluation.

\subsubsection{LLMs in HalluJudge}
To explore the assessment strategies in HalluJudge spanning direct assessment, few‑shot direct assessment, multi‑step reasoning, and tree‑of‑thoughts reasoning, we require LLMs that support long‑context analysis, execute explicit multi‑hop or tree‑structured reasoning, and adaptable computation for this task complexity. 
We therefore employ Gemini 3 and GPT‑5.1. 
Gemini 3 is a latest LLM, designed for state‑of‑the‑art reasoning and nuanced problem decomposition with context windows up to 1 million tokens, making it suitable for reasoning chains and tree‑structured deliberation over long artifacts.
In this work, we use \texttt{Gemini-3-pro-preview} to infer the judgment. 
GPT‑5.1 is another recent LLM with strong adaptive reasoning capability.
It dynamically allocates ``thinking time'' to complex reasoning prompts and tuned for multi‑step problem solving and instruction following; the API also exposes configurable reasoning effort and large context (up to ~400k tokens), aligning with our need to compare direct, few‑shot, and tree‑of‑thoughts regimes under consistent conditions.  
In this work, we use \texttt{GPT-5.1-2025-11-13} to infer the judgment.

\subsubsection{Analysis Approach and Evaluation}
To answer our RQs, we evaluate the performance of four assessment strategies and two LLMs within  HalluJudge. Our evaluation proceeds as follows.

For \textbf{RQ1}, we quantify the effectiveness of hallucination detection by measuring Precision ($\frac{TP}{TP + FP}$), Recall ($\frac{TP}{TP + FN}$), and the F-measure ($\frac{2 \times \text{Precision} \times \text{Recall}}{\text{Precision} + \text{Recall}}$), where a true positive (TP) indicates a case in which HalluJudge correctly identifies a hallucinated review comment that was annotated as hallucinated by human annotators. False positives (FP) and false negatives (FN) correspond to incorrect hallucination judgments relative to human annotations.

For \textbf{RQ2}, we quantify efficiency along two dimensions: (i) the size of the inputs and outputs, and (ii) the estimated monetary cost. 
We use the number of used tokens (i.e., token count) as a proxy for computation time because the number of tokens strongly correlates with LLM inference effort, while avoiding external variability such as network delays.
The input consists of the system prompt and user prompt, while the output comprises the model's answer and explanation. 
To ensure accurate and model‑specific token count, we use the official tokenizer APIs provided by each LLM.\footnote{\url{https://ai.google.dev/gemini-api/docs/tokens?lang=python} and \url{https://platform.openai.com/tokenizer}}
For each hallucination assessment, we estimate the monetary cost as a linear function of the input and output token counts $Cost= c_{in}\times \mathrm{InputTokens} + c_{out} \times \mathrm{OutputTokens}$.
We use the provider's standard pricing at evaluation time for the estimation, i.e., Gemini 3 is priced at \$2.00 per 1M input tokens and \$12.00 per 1M output tokens, while GPT‑5.1 is priced at \$1.25 per 1M input tokens and \$10.00 per 1M output tokens.

For \textbf{RQ3}, we quantified the alignment between HulluJudge's assessment and developer preferences by examining two key dimensions: consistency and preference coverage.
\textbf{Consistency} measures the proportion of generated review comments that are both judged as non‑hallucinated ($H'$) by HalluJudge and received a thumbs‑up ($P$) from developers, relative to all comments judged as non‑hallucinated ($H'$), i.e., $\frac{H' \cap P}{H'}$.
A high consistency score indicates that HalluJudge’s assessment are well aligned with what developers actually prefer.
\textbf{Coverage} measures the proportion of generated comments that are both judged as non‑hallucinated ($H'$) and receive a thumbs‑up ($P$), relative to all comments that receive a thumbs‑up from developers ($P$), i.e., $\frac{H' \cap P}{P}$.
A high coverage score indicates that HalluJudge successfully captures the majority of comments that developers consider useful.



\section{Experimental Results}
We now present the results of our three research questions.

\input{tables/accuracy_only_result}

\subsection*{RQ1: \rqone}
\noindent \underline{\textbf{Results.}}
Table~\ref{table:concise_table} shows the results of hallucination detection for the various prompt and LLMs in HalluJudge.
Overall, Gemini 3 with the tree of thought assessment achieves the strongest performance, consistently reaching 0.85 for precision, recall, and F1.
We also observe that tree of thought is consistently the top performing assessment strategy among the four strategies, followed by basic-zero-shot as the second best performing strategy.
For both LLMs, multi-step-reasoning and few-shot achieved lower performance.
The relative ranking of the four strategies remains stable across both  Gemini 3 and GPT 5.1.
These results suggest that defining explicit reasoning structures
helps an LLM to assess how well a review comment is grounded in its corresponding code diff.

We find that Gemini 3 achieves relatively higher performance  than GPT 5.1.
Specifically, the least effective strategies for Gemini 3 (i.e., few shot and multi-step-reasoning achieved F1 scores of 0.79.
In contrast, the most effective strategy for GPT 5.1 (i.e., tree of thought) achieves an F1 score of 0.76.
In terms of the sensitivity, Gemini 3 exhibits less variation in F1 score than GPT 5.1, with the former varying by at most 0.06 and the latter by at most 0.09.
In both cases, we find that the fluctuation can be mostly attributed to a variation in recall, rather than precision.
For Gemini 3, precision and recall can vary by as much as 0.04 and 0.07, respectively.
For GPT 5.1, precision and recall can vary by as much as 0.04 and 0.10, respectively.

\begin{summarybox}[colback=blue!3,colframe=blue!50]{RQ1 Summary}
HalluJudge effectively detects hallucinations in code review comments, achieving a precision, recall, and F1 score of 0.85. Among the four assessment strategies, the tree of thought approach delivers the highest scores across all three metrics.
\end{summarybox}

\begin{figure}[t]
    \centering
    \includegraphics[width=\columnwidth, trim={0 25 0 0}]{figures/combined.png}
    \caption{(RQ2) The distribution of token counts and monetary costs for each inference of hallucination judgment.}
        \label{fig:combined}
\end{figure}
\subsection*{RQ2: \rqtwo}
\noindent \underline{\textbf{Results.}}
To compare the efficiency of the hallucination judges in a standardized manner, we analyze the input/output tokens incurred by the four strategies and LLMs.
Figure \ref{fig:combined}  shows the distributions of the input tokens, output tokens, and cost.\footnote{The same system prompt results in a different token count due to model-specific tokenizers.}

In general, the direct assessment strategy uses the fewest tokens and therefore incurs the lowest cost, despite being the second‑best performing prompt.
On average, each hallucination assesment of a generated review comment requires 2,784–3,315 input tokens and 81–89 output tokens.
This results in an average cost of \$0.009  per inference for Gemini 3 and \$0.004 per inference for GPT‑5.1.
On the other hand, the tree of thought strategy which is the best performer requires substantially more input and output tokens, resulting in an additional cost of \$0.005 for Gemini 3 and \$0.007 for GPT 5.1 per inference.
We also find that both few shot and multi-step reasoning incurs higher cost and token counts than direct assessment, despite achieving lower performance.
Whilst the few shot strategy only costs up to \$0.001  more on average than the direct assessment, the multi-step reasoning strategy can cost up to \$0.003 more on average. 
These results suggest that direct assessment is the most cost‑effective strategy when considering both hallucination‑detection performance and computational cost.

The discrepancy of token counts and cost is largely due to the  length of the assessment instructions in the system prompt and the average length of the elicited output. 
Specifically, the system prompt of direct assessment contains 313 tokens for Gemini 3 and 302 tokens for GPT 5.1, whilst the system prompt of the tree of thought strategy requires an additional of 534 and 508 tokens, respectively.
For the output token, direct assessment produces short answers and explanation, yielding outputs with an average of 89 and 81 tokens for Gemini 3 and GPT 5.1, respectively.
In contrast,  the tree of thought strategy elicits the explanation for each reasoning branch, costing an additional 418 and 577 output tokens on average.
The multi-step-reasoning can elicit up to 189 more tokens on average than the direct assessment, whilst the few shot only elicits up to 6 more tokens on average.
Comparing across models, GPT 5.1's tokenizer can save up to 56 input tokens depending on the system prompt, but may also elicit up to 151 more output tokens on average, as seen in the case of tree-of-thought.

\begin{summarybox}[colback=blue!3,colframe=blue!50]{RQ2 Summary}
The direct assessment in HalluJudge is cost-effective in terms of the token used and monetary cost, whereas the tree of thought strategy achieves the best performance but requires the highest cost.
\end{summarybox}

\subsection*{RQ3: \rqthree}
\noindent \underline{\textbf{Results.}}
To further validate our human annotation results, we measure the alignment between  HulluJudge’s assessment and developer preference of the actual review comments generated by the automated code reviewer (i.e., RovoDev Code Reviewer) during the online deployment.
Specifically, we examine the alignment of HulluJudge with the thumb-up reaction which indicates an approval of a comment by developers.

Table~\ref{table:online_dataset} presents the result of the alignment between HalluJudge's judgments and developer preferences in terms of consistency and coverage. 
The last column provides the average of the consistency and coverage values.
In general, we find that HalluJudge aligns well with developer preferences. 
It achieves a consistency score of 0.67-0.72 and a coverage score of 0.53-0.65. 
These results indicate that up to 72\% of comments classified by HalluJudge as non‑hallucinated also receive a thumbs‑up reaction from developers, and that HalluJudge successfully captures up to 65\% of all comments that receive a thumbs‑up.
These results suggest that comments judged as non‑hallucinated are likely to receive developer approval.

Consistent with the RQ1 results based on the human annotation dataset, we find that the previous top assessment strategies (i.e., tree of thought and direct assessment with Gemini 3) also achieve the high consistency and coverage with an average of of 0.66 and 0.67, respectively.
Overall, we observe the consistent ranking in terms of  effectiveness results in RQ1 and alignment results in RQ3, highlighting that the assessment strategies most effective at detecting hallucinations are also the ones that best reflect developer preferences in real‑world use.

Comparing between the models, we find that Gemini 3 still outperforms GPT 5.1, regardless of assessment strategies.
Specifically, the least effective strategy (i.e., few shot)  with Gemini 3, achieved an average of 0.65, whilst multi-step-reasoning with GPT 5.1 could only achieve an average of 0.63.
This is driven by substantially lower coverage for the various strategies with  GPT 5.1

\begin{summarybox}[colback=blue!3,colframe=blue!50]{RQ3 Summary}
HalluJudge assessment is well aligned with the developer preference signals (i.e., Thumbs Up) in the online production with an average consistency and coverage score up to 0.67.
Consistent with the human annotation dataset, the tree of thought and direct assessment with Gemini 3 remain top-performing.
\end{summarybox}

\input{tables/online_data}

\begin{figure}[t]
  \centering
  \small
  
  \begin{tabular}{p{0.98\linewidth}}
    \\ \hline
    \textbf{Diff (Java):} \\
    \texttt{@@ -42,1 +42,6 @@ public final class ExampleContext \{\}} \\
    \texttt{    private Optional<String> getAuthorizationToken() \{} \\
    \texttt{+        logger.info(} \\
    \texttt{+            (requestContext.getHeader(AUTH\_HEADER) != null)} \\
    \texttt{+                ? "auth-header-length = " + requestContext.getHeader(AUTH\_HEADER).length()} \\
    \texttt{+                : "auth-header not found"} \\
    \texttt{+        );} \\
    \hline
    \textbf{LLM Code Review Comment (Hallucinated):} \\
    \emph{``Consider using isPresent() on the Optional returned by getHeader instead of checking for null to maintain consistency with Optional usage patterns.''} \\
    \hline
    \textbf{Why this is hallucination? (Human Judge):} \\
    This comment assumes \texttt{requestContext.getHeader(AUTH\_HEADER)} returns an \texttt{Optional<String>}, when in reality (as evidenced by length()) it returns a plain String, which does not have an \texttt{isPresent()} method.\\
    \hline
    \textbf{HalluJudge's Direct Assessment:} \\
    \textcolor{purple}{\textbf{Answer: } 4.}
    Completely misaligned because \textcolor{ForestGreen}{it incorrectly assumes that \texttt{requestContext.getHeader()} returns an Optional}. \textcolor{RoyalBlue}{The code diff shows \texttt{.length()} being called directly on the return value}, 
    which indicates the return type is a String.
    Consequently, the suggestion to use \texttt{isPresent()} is technically invalid for the actual data type. \\
    \hline
    \textbf{Tree of Thought:} \\
    \textcolor{purple}{\textbf{Answer:} Context\_Misalignment.}
    The comment is factually incorrect regarding the return type of \texttt{getHeader}. The code diff shows \texttt{.length()} being called directly on \texttt{requestContext.getHeader(AUTH\_HEADER)}. Since \texttt{java.util.Optional} does not have a \texttt{.length()} method, \texttt{getHeader} must return a String, not an Optional. Therefore, the suggestion to use \texttt{isPresent()} is syntactically invalid and contradicts the code evidence. \\
    \textcolor{purple}{\textbf{Log: }}[\{
    \textcolor{ForestGreen}{\textbf{branch\_1\_claims}: getHeader returns an Optional.} 
    Verification: The code executes .length(). Optional does not possess a \texttt{.length()} method. Conclusion: The claim is false..., \textbf{branch\_2\_patterns:}  Adding a debug log statement. Comment Pattern: Code style/API usage suggestion. Appropriateness: The suggestion is technically inapplicable ...,
    \textcolor{RoyalBlue}{\textbf{branch\_3\_coverage}: The comment targets the added logging line. However, it fails to account for the \texttt{.length()} method call}..., 
    \textcolor{ForestGreen}{\textbf{branch\_4\_assumptions:} The LLM assumes \texttt{requestContext.getHeader()} follows a pattern where it returns Optional. This is a common pattern in some modern libraries but contradicts the standard Servlet API (which returns String)} ..., \textcolor{ForestGreen}{\textbf{branch\_5\_alternatives:} It is possible the LLM confused the return type of the enclosing method (\texttt{Optional<String> getAuthorizationToken})...,} \textbf{Synthesis:} The comment is based on a hallucinated API signature. The code usage \texttt{.length()} proves the object is not an Optional... The evidence for misalignment is strong and unambiguous., \textcolor{purple}{\textbf{Answer:} 4}, \textcolor{purple}{\textbf{Confidence:} High}, Evidence\_Summary: The code calls \texttt{.length()} on the return value of getHeader, proving ... which would cause a compilation error.\}]\\
    \hline
  \end{tabular}
  \caption{An example of HalluJudge's explanation of a hallucinated code review comment.}
  \label{fig:case_study}
\end{figure}


\section{Discussion \& Implications}

We now further discuss the experimental results and provide implications of our findings.

\underline{\textbf{Discussion 1:}} \emph{Do the diverse assessment strategies provide complementary judgment?}
In this work, we evaluated each of HalluJudge’s assessment strategies independently.
However, it is plausible that aggregating the judgments from all four strategies could yield stronger overall performance.
To investigate this possibility, we further examined the effectiveness of an ensemble approach.
Following prior work~\cite{zhou2025llm}, we compute the average score across the four assessment strategies and use this aggregated score to determine whether a review comment is hallucinated.
We consider a comment to be hallucinated when its average score exceeds 0.5, i.e.,  at least two strategies assign a score greater than zero.
We find that the aggregated assessment using Gemini 3 yields a precision of 0.83, a recall of 0.80, and an F1‑score of 0.81.
Similarly, the aggregated assessment using GPT‑5.1 attains a precision of 0.79, a recall of 0.71, and an F1‑score of 0.73.
However, these results are lower than those achieved by the the tree of thought assesment alone, which achieves an F1‑score of 0.85 (Gemini 3) and 0.79 (GPT‑5.1).
This suggests that the assessment strategies do not provide complementary signals, and combining them does not improve overall effectiveness.

\underline{\textbf{Discussion 2:}} \emph{Is HalluJudge's assessment based on the degree to which a review comment is grounded in its context?}
Since HalluJudge is explicitly instructed to generate an explanation, allowing us to inspect its decision rationale. This transparency helps us confirm correct reasoning of the assessment.
To demonstrate the reasoning process of HalluJudge, 
Figure~\ref{fig:case_study} shows examples of HalluJudge's responses and explanations based on the two top effective strategies (i.e., Direct and Tree of Thought).
In this scenario, the change introduces logic that checks if the authorization header exists then logs the character count (\texttt{.length()}), otherwise logs a "not found" message.
The LLM review comment incorrectly suggests that  the return value of the authorization header should be an \texttt{Optional<String>} rather than a plain string, and proposes an alternative implementation to check existence with \texttt{isPresent()}, a standard, built-in method of the java.util.Optional class.
As the actual type of the authorization header is \texttt{String}, the built-in method does not exist.

In this scenario, both direct  and tree of thought assessments correctly identify the incorrect assumption that the authorization header returns an \texttt{Optional<String>} (highlighted in green). 
By citing the use of length() as evidence in code diff (highlighted in blue), they correctly deduce that the call to \texttt{isPresent()} is syntactically invalid.
These explanations demonstrate the effectiveness of evidence-based reasoning anchored in the code diff context, which enabled HalluJudge to pinpoint factual contradictions and accurately assign the score of 4 (Completely Misaligned) to the comment.
In addition, tree of thought imposes a branching structure, providing an extended reasoning log that offers a more in-depth analysis of the scenario.
Branches 4 and 5 extend the analysis by investigating causality, a step missing from the direct assessment. They attribute the error to potential bias from common API patterns (parametric knowledge) or a more probable reason: confusion with the enclosing method's \texttt{Optional<String>} signature.
We observe that the explanations of both the direct and the tree‑of‑thought assessments are highly consistent across the human annotation dataset. 
In addition, tree‑of‑thought approach provides an additional advantage by offering more elaborated reasoning through its branching structure. 
This observation suggest that both assessment strategies offer reliable and explainable judgment of hallucination.


\underline{\textbf{Implication 1:}} \emph{HalluJudge can provide a practical safeguard mechanism for detecting context-misaligned code review comments before they reach developers.} 
RQ1 demonstrates that HalluJudge, particularly when using the direct and tree of thought assessments can effectively detect hallucinated review comments.
RQ3 further shows that HalluJudge’s judgments are well aligned with developer preferences of actual LLM review comments in production.
Together, these findings suggest that explicitly reasoning about claim grounding, traceability, and scope boundaries between a generated review comment and its  context (i.e., the code diff) is a promising approach for identifying and mitigating hallucinations in code review.
Integrating HalluJudge as a filtering or validation layer could reduce developers’ exposure to hallucinated comments, addressing a key barrier to the adoption of LLM‑based code reviewers.
By preventing plausible yet incorrect feedback from reaching users, such integration may enhance perceived reliability and foster long‑term trust in AI‑assisted review tools.

\underline{\textbf{Implication 2:}} \emph{HalluJudge provides a cost‑effective and explainable approach to hallucination detection in code reviews.}
RQ2 shows that the cost of detecting hallucinations is relatively low, for example, direct assessment with Gemini 3 requires on average only \$0.009 per judgment.
This suggests that HalluJudge is a cost-effective alternative to manual evaluation with experienced human annotators, which has previously been the sole reliable standard for evaluating the correctness of LLM-generated code review comments~\cite{lin2024leveraging,tufano2024code,learning_from_experience}.
Finally, our observation in Discussion 2 show that HalluJudge’s explanations provide clear and interpretable reasoning behind its judgments.
These informative explanations can support practitioners and tool builders in understanding why a comment is considered hallucinated, further enhancing transparency in the assessment process.




\section{Threats to the Validity}
 We discuss potential threats to the validity of our results and discuss the steps taken to mitigate them where possible.

\textbf{Internal validity.} 
The construction of ground-truth labels for hallucination in RQ1 may be subject to individual bias or inconsistency in interpreting context misalignment. To mitigate this, annotations followed a clearly defined operational definition of hallucination grounded in traceability to the code diff and adherence to change intent. The annotation process required explicit verification of factual support, reducing reliance on subjective judgment.



\textbf{Construct validity.} 
In this work, hallucination is operationalized as context misalignment, i.e., claims in a review comment that are unsupported by, contradictory to, or outside the scope of the code diff. While this definition captures a critical and practically harmful class of hallucinations, it may not cover all possible quality issues in code review comments, such as stylistic redundancy or lack of actionable detail.

For RQ3, we use developer thumbs-up and thumbs-down feedback as a proxy for review comment quality. However, developers may provide a thump up reaction for various reasons (e.g., politeness) or a thumb down due to personal preference. However, prior work suggests that implicit developer feedback provides ecologically valid signals of usefulness in production settings~\cite{Tantithamthavorn2026RovoDevCR}, and our goal in RQ3 is to measure the alignment with developer preferences rather than absolute correctness.

\textbf{External validity.} 
Our datasets are drawn from Atlassian’s enterprise-scale software projects, which may differ from open-source or smaller industrial environments in terms of codebase size, review practices, and available context. As a result, the prevalence and nature of hallucinations, and the effectiveness of HalluJudge may differ in other settings.
We partially mitigate this threat by evaluating HalluJudge across a diverse set of repositories, programming languages, and projects, and by using multiple underlying LLMs. Nevertheless, 
future work is needed to replicate our findings in open-source ecosystems and alternative review workflows.

\section{Conclusions}

LLM-generated review comments can be hallucinated, i.e., ungrounded in the actual code---poses a significant challenge to the adoption of LLMs in code review workflows. 
In this work, we explore effective and scalable methods for the hallucination detection without the reference.  
We design HalluJudge that includes four key strategies to assess the grounding of generated review comments based on the context alignment. 
Based on Atlassian's enterprise-scale software projects, our results show that the hallucination assessment in HalluJudge is cost-effective with an F1 score of 0.85 and an average cost of \$0.009. 
On average, 67\% of the HalluJudge assessments are aligned with the developer preference of the actual LLM-generated review comments in the online production.
Our results suggest that HalluJudge can serve as a practical
safeguard to reduce developers’ exposure to hallucinated comments,
fostering trust in AI-assisted code reviews.

\bibliographystyle{ACM-Reference-Format}

\bibliography{references}
\balance
\end{document}

%% file: tables/annotation_dataset.tex
\begin{table}[t]
\caption{A summary of our human-annotated dataset.}
\centering
\footnotesize
\begin{tabular}{llcc}
\toprule
\rowcolor[HTML]{FFFFFF} 
Stratum                                                                                                                          & \multicolumn{1}{c}{\cellcolor[HTML]{FFFFFF}Group} & \#Comments & \% of Total \\ \hline

\rowcolor[HTML]{EFEFEF} 

\multicolumn{1}{l|}{\cellcolor[HTML]{EFEFEF}}                                                                                    & Java                                              & 58        & 40\%        \\
\rowcolor[HTML]{FFFFFF} 
\multicolumn{1}{l|}{\cellcolor[HTML]{EFEFEF}}                                                                                    & Python                                            & 16        & 11\%        \\
\rowcolor[HTML]{EFEFEF} 
\multicolumn{1}{l|}{\cellcolor[HTML]{EFEFEF}}                                                                                    & JavaScript                                        & 21        & 15\%        \\
\multicolumn{1}{l|}{\cellcolor[HTML]{EFEFEF}}                                                                                    & TypeScript                                        & 24        & 17\%        \\
\rowcolor[HTML]{EFEFEF} 
\multicolumn{1}{l|}{\multirow{-5}{*}{\cellcolor[HTML]{EFEFEF}Language}}                                                          & Kotlin                                            & 24        & 17\%        \\ \hline
\rowcolor[HTML]{FFFFFF} 
\multicolumn{1}{l|}{\cellcolor[HTML]{EFEFEF}}                                                                                    & Add (+)                                           & 74        & 52\%        \\
\rowcolor[HTML]{EFEFEF} 
\multicolumn{1}{l|}{\multirow{-2}{*}{\cellcolor[HTML]{EFEFEF}Change Type}}                                                       & Modify (+-)                                       & 69        & 48\%        \\ \hline
\rowcolor[HTML]{FFFFFF} 
\multicolumn{1}{l|}{\cellcolor[HTML]{EFEFEF}}                                                                                    & Context Misaligned                            & 32        & 22\%        \\             
\rowcolor[HTML]{EFEFEF} 
\multicolumn{1}{l|}{\multirow{-2}{*}{\cellcolor[HTML]{EFEFEF}\begin{tabular}[c]{@{}l@{}}Hallucination\end{tabular}}} & Context Aligned                                         & 111        & 78\%        \\ \hline

\label{table:offline_dataset}
\end{tabular}
\end{table}

%% file: tables/accuracy_only_result.tex
\begin{table}[t]
\caption{(RQ1) The effectiveness of different assessment strategies of Hallucination Judges. Bold text represents \#1, and underlined text represents \#2.}
\centering
\footnotesize

\begin{tabular}{lllll}
\toprule
\multicolumn{1}{c}{Assessment Strategies} & \multicolumn{1}{c}{LLM}                       & \multicolumn{1}{c}{Precision} & \multicolumn{1}{c}{Recall} & \multicolumn{1}{c}{F1}                                \\ \hline
\rowcolor[HTML]{EFEFEF} 
Tree of Thought                           & \multicolumn{1}{l}{\cellcolor[HTML]{EFEFEF}Gemini 3} & \textbf{0.85}         & \textbf{0.85}         & \multicolumn{1}{l}{\cellcolor[HTML]{EFEFEF}\textbf{0.85}}      \\
\rowcolor[HTML]{FFFFFF} 
Direct                           & \multicolumn{1}{l}{\cellcolor[HTML]{FFFFFF}Gemini 3} & \ul{ 0.83}            & \ul{ 0.81}            & \multicolumn{1}{l}{\cellcolor[HTML]{FFFFFF}\ul{ 0.82}}  \\

\rowcolor[HTML]{EFEFEF} 
Few-Shot                                   & \multicolumn{1}{l}{\cellcolor[HTML]{EFEFEF}Gemini 3} & 0.81                  & 0.78                  & \multicolumn{1}{l}{\cellcolor[HTML]{EFEFEF}0.79}    \\
\rowcolor[HTML]{FFFFFF} 
Multi-Step Reasoning                      & \multicolumn{1}{l}{\cellcolor[HTML]{FFFFFF}Gemini 3} & 0.81                  & 0.78                  & \multicolumn{1}{l}{\cellcolor[HTML]{FFFFFF}0.79}    \\
\rowcolor[HTML]{EFEFEF} 
Tree of Thought                           & \multicolumn{1}{l}{\cellcolor[HTML]{EFEFEF}GPT 5.1}   & 0.79                  & 0.74                  & \multicolumn{1}{l}{\cellcolor[HTML]{EFEFEF}0.76}    \\
\rowcolor[HTML]{FFFFFF} 
Direct                          & \multicolumn{1}{l}{\cellcolor[HTML]{FFFFFF}GPT 5.1}   & 0.77                  & 0.74                  & \multicolumn{1}{l}{\cellcolor[HTML]{FFFFFF}0.75}         \\
\rowcolor[HTML]{EFEFEF} 
Multi-Step Reasoning                      & \multicolumn{1}{l}{\cellcolor[HTML]{EFEFEF}GPT 5.1}   & 0.77                  & 0.73                  & \multicolumn{1}{l}{\cellcolor[HTML]{EFEFEF}0.74}       \\
\rowcolor[HTML]{FFFFFF} 
Few-Shot                                   & \multicolumn{1}{l}{\cellcolor[HTML]{FFFFFF}GPT 5.1}   & 0.75                  & 0.64                  & \multicolumn{1}{l}{\cellcolor[HTML]{FFFFFF}0.67}      \\ \hline


\label{table:concise_table}
\end{tabular}

\end{table}

%% file: tables/online_data.tex
\begin{table}[t]
\caption{(RQ3) An evaluation of the alignment (i.e., consistency and coverage) between HalluJudge's assessment and developer preferences. Bold text represents \#1, and underlined text represents \#2.}
\centering
\footnotesize
\begin{tabular}{lllll}
\toprule
\multicolumn{1}{c}{}                        & \multicolumn{1}{c}{}                                              & \multicolumn{3}{c}{Developer Preference}                                          \\
\multicolumn{1}{c}{\multirow{-2}{*}{Judge}} & \multicolumn{1}{c}{\multirow{-2}{*}{Model}}                       & \multicolumn{1}{c}{Consistency} & \multicolumn{1}{c}{Coverage} & \multicolumn{1}{c}{Average} \\ \hline
\rowcolor[HTML]{EFEFEF} 
Tree of Thought                           & \multicolumn{1}{l|}{\cellcolor[HTML]{EFEFEF}Gemini 3} & 0.67                  & \textbf{0.65}         & \ul{ 0.66}             \\
\rowcolor[HTML]{FFFFFF} 
Direct                           & \multicolumn{1}{l|}{\cellcolor[HTML]{FFFFFF}Gemini 3} & 0.69                  & \textbf{0.65}         & \textbf{0.67}          \\
\rowcolor[HTML]{EFEFEF} 
Few Shot                                  & \multicolumn{1}{l|}{\cellcolor[HTML]{EFEFEF}Gemini 3} & 0.67                  & 0.62                  & 0.64                   \\
\rowcolor[HTML]{FFFFFF} 
Multi-Step Reasoning                      & \multicolumn{1}{l|}{\cellcolor[HTML]{FFFFFF}Gemini 3} & 0.68                  & \ul{ 0.63}            & 0.65                   \\
\rowcolor[HTML]{EFEFEF} 
Tree of Thought                           & \multicolumn{1}{l|}{\cellcolor[HTML]{EFEFEF}GPT-5}   & \textbf{0.72}         & 0.54                  & 0.61                   \\
\rowcolor[HTML]{FFFFFF} 
Direct                           & \multicolumn{1}{l|}{\cellcolor[HTML]{FFFFFF}GPT-5}   & 0.69                  & 0.54                  & 0.61                   \\
\rowcolor[HTML]{EFEFEF} 
Multi-Step Reasoning                      & \multicolumn{1}{l|}{\cellcolor[HTML]{EFEFEF}GPT-5}   & \ul{ 0.70}            & 0.57                  & 0.63                   \\
\rowcolor[HTML]{FFFFFF} 
Few Shot                                  & \multicolumn{1}{l|}{\cellcolor[HTML]{FFFFFF}GPT-5}   & 0.69                  & 0.53                  & 0.60                   \\ \hline


\label{table:online_dataset}
\end{tabular}
\end{table}